\documentclass[12pt, draftclsnofoot, onecolumn]{IEEEtran}

\usepackage{amssymb}
\usepackage{amsmath}
\usepackage{cite}
\usepackage{url}
\usepackage{xcolor}
\usepackage{cite,graphicx,amsmath,amssymb}
\usepackage{subfigure}
\usepackage{citesort}
\usepackage{fancyhdr}
\usepackage{mdwmath}
\usepackage{mdwtab}
\usepackage{caption}
\usepackage{amsthm}
\usepackage{algorithm}
\usepackage{algorithmic}
\usepackage{graphicx} 
\usepackage{epstopdf}
\usepackage{setspace}
\usepackage{tikz}
\usepackage{amsmath, bm}
\usepackage{caption}
\usepackage[justification=centering]{caption}

\newtheorem{remark}{Remark}
\newtheorem{theorem}{Theorem}

\newtheorem{lemma}{Lemma}

\newtheorem{corollary}{Corollary}

\makeatletter
\def\ScaleIfNeeded{%
\ifdim\Gin@nat@width>\linewidth \linewidth \else \Gin@nat@width
\fi } \makeatother

\begin{document}

\title{\huge{Mobile Reconfigurable Intelligent Surfaces for NOMA Networks: Federated Learning Approaches}
}

\author{\normalsize {Ruikang~Zhong,~\IEEEmembership{\normalsize Student Member,~IEEE,}
Xiao~Liu,~\IEEEmembership{\normalsize Student Member,~IEEE,}
Yuanwei~Liu,~\IEEEmembership{\normalsize Senior Member,~IEEE,}
Yue~Chen,~\IEEEmembership{\normalsize Senior Member,~IEEE,}
Zhu~Han,~\IEEEmembership{\normalsize Fellow,~IEEE}}

\thanks{Ruikang~Zhong, Xiao~Liu, Yuanwei~Liu, and Yue~Chen are with the Queen Mary University of London, London E1 4NS, U.K. (e-mail: r.zhong@qmul.ac.uk; x.liu@qmul.ac.uk; yuanwei.liu@qmul.ac.uk; yue.chen@qmul.ac.uk).
Zhu~Han is with the Department of Electrical and Computer Engineering, University of Houston, Houston, TX 77004, USA (e-mail: zhan2@uh.edu).
}
}

\maketitle
\begin{abstract}
A novel framework of reconfigurable intelligent surfaces (RISs)-enhanced indoor wireless networks is proposed, where an RIS mounted on the robot is invoked to enable mobility of the RIS and enhance the service quality for mobile users. Meanwhile, non-orthogonal multiple access (NOMA) techniques are adopted to further increase the spectrum efficiency since RISs are capable to provide NOMA with artificial controlled channel conditions, which can be seen as a beneficial operation condition to obtain NOMA gains. To optimize the sum rate of all users, a deep deterministic policy gradient (DDPG) algorithm is invoked to optimize the deployment and phase shifts of the mobile RIS as well as the power allocation policy. In order to improve the efficiency and effectiveness of agent training for the DDPG agents, a federated learning (FL) concept is adopted to enable multiple agents to simultaneously explore similar environments and exchange experiences. We also proved that with the same random exploring policy, the FL armed deep reinforcement learning (DRL) agents can theoretically obtain a reward gain compare to the independent agents.
Our simulation results indicate that the mobile RIS scheme can significantly outperform the fixed RIS paradigm, which provides about three times data rate gain compare to the fixed RIS paradigm. Moreover, the NOMA scheme is capable to achieve a gain of 42\% in contrast with the OMA scheme in terms of sum rate. Finally, the multi-cell simulation proved that the FL enhanced DDPG algorithm has a superior convergence rate and optimization performance than the independent training framework.


\end{abstract}


\section{Introduction}

Reconfigurable intelligent surfaces (RISs) \cite{9133157}, also known as intelligent reflecting surfaces (IRSs) \cite{wu2019towards}, have been anticipated as a neonatal component of future communication systems \cite{8869705}. By employing a number of arranged reflecting elements, the signal can be reflected by RISs to provide additional channels for wireless links \cite{9139273,9133094}. Therefore, once the phases of reflection elements are coordinated in a well-organized manner, an effect of passive beamforming \cite{9174801} can be achieved and the reflected signal can be concentrated on users to provide considerable channel gains \cite{9086766}. One of the main factors that RISs can provide noticeable gain is that they can provide further possible line-of-sight (LoS) paths for users who do not originally have an LoS path \cite{han2019large}. However, in most existing research contributions, RISs are fixed on a wall or other bearing, and therefore the fixed deployment causes RISs may not be able to obtain LoS paths and optimal channel enhancement, especially in the environment with obstructions.
In an effort to complement this defect, in this paper, we propose a mobile RIS model that RISs are mounted on intelligent robots to achieve its flexible deployment.

Another compelling concern in the communication field is the user capacity since the number of users brought by the Internet of Things (IoT) is upstaging continuously \cite{shirvanimoghaddam2017massive}. As a consequence, to further improve the capacity and spectrum efficiency of wireless networks, non-orthogonal multiple access (NOMA) techniques have become a highly sought-after candidate technique \cite{yang2016general}. Moreover, NOMA techniques have been proved to be capable of achieving several advantages specifically in the RIS-assisted communication network. As pointed out by the authors of \cite{liu2020reconfigurable2}, affinities between RISs and the NOMA scheme include that RISs can provide additional signal diversity, desired channel condition, and undemanding multi-antenna constrain for NOMA systems. The main interplay is that in a conventional NOMA enhanced wireless network, the decoding order of successive interference cancelation (SIC) determined by the natural channel state information (CSI) of users, which is not likely to be fully consistent with the users' data rate demand. However, RISs can artificially modify the CSI for each user and thereby provide desired propagation condition for superposed signals. Therefore, NOMA techniques are invoked in our mobile RIS model to obtain further capacity and data rate gains.

To maximize the profit of empowering mobility to RISs, how to plan proper dynamic deployments for mobile RISs are a problem worth exploring. Since users are considered as moving as well, the optimization problem is highly dynamic, and the joint optimization problem of movements and phase shifts of RISs is an emerging problem worth exploring. In addition, since obstacles that hinder the movement of RISs and shields LoS paths are likely to have irregular and non-analytic shapes, this also raises challenges for conventional optimization approaches.
In contrast to convex optimization, deep reinforcement learning (DRL) is considered to be a more competent methodology for dynamic optimization problems since DRL is able to recognize the current state of the environment~\cite{chen2019artificial,6542770}. Meanwhile, since multiple mobile RISs can be deployed in different cells, federated learning (FL) is employed to strengthen their training efficiency and effectiveness for the proposed multi-cell multi-agent scenario \cite{li2020federated}. FL arouses the interest of researchers as a distributed learning framework since it can effectively utilize computational resources \cite{kang2020reliable} with a protection of user privacy \cite{park2020communication}. Especially for the DRL algorithm, FL can improve training efficiency and learning effect, since agents can explore the environments simultaneously and their knowledge can be transferred to each other through a global neural networks model.
Therefore, we propose a DRL algorithm with a framework of FL, namely the FL enhanced deep deterministic policy gradient (FL-DDPG) algorithm to jointly optimize the passive beamforming, dynamic deployment of RISs, and the power allocation for NOMA users.


Although the enthusiasm of RISs and machine learning in recent years has resulted in that a number of related researches have been completed, distinguished from the existing research contributions, we propose the following new contributions.

\begin{itemize}

\item We propose a novel indoor communication model, which employs mobile RIS to enhance the channel quality for users. Compared to existing fixed RIS paradigms, the proposed framework is capable to cover indoor users who suffered from obstructed environments with the aid of flexible deployments of RISs, thereby increasing the sum data rate. In order to further increase user capacity and increase spectrum efficiency, NOMA techniques are invoked. A corresponding dynamic decoding order scheme is adopted, since the channels intervened by mobile RISs are likely to significantly impact the user's CSI. Build on the proposed mobile RIS framework, we formulate the maximization problem of the sum data rate.

\item We invoke the DDPG algorithm to jointly optimize the deployments, phase shifts of mobile RISs and the power allocation policy for users. Since the fading matrixes of users are input into the neural network, the dimension of the input state can have a significant difference. Therefore, the size of the neural network has to be adaptive accordingly to ensure an effective and precise fitting and give the corresponding empirical formula.

\item We propose a federated learning enabled DRL framework to reduce the training time of agents and theoretically prove that within limited training processes, the FL framework is capable to provide reward gains for DRL agents. We invoke an FL model with local training and periodic global model update to enable the agent in each cell to learn from others' experiences and thereby improve the efficiency of exploration and training. In addition, we also investigate the impact of the different propagation characteristics of each cell on FL learning effects.
    Our simulation results proved that with finite training episodes, the DRL algorithm enhanced by FL is capable to obtain superior performance compared to the independent agent training approach.

\end{itemize}

%
%
%
%
%


Section \ref{section:2} reviews related state of the art contributions. Section \ref{section:3} illustrates the mobile RISs aided indoor communication models, including both OMA and NOMA scheme and the problem formulation. Section \ref{section:4} introduces the FL framework, which is employed to coordinate multi-cell optimization. Section  \ref{section:5} presents the FL-DDPG algorithm for the joint optimization of the user power allocation policy, the deployment and phase shift of the mobile RIS. Section  \ref{section:6} demonstrates and analyzes the simulation results. Finally, Section \ref{section:7} is the conclusion section of this paper.

\section{The State of the Art}\label{section:2}

This section briefly reviews the state of the art research on the RISs assisted NOMA network and DRL/FL optimization in wireless network fields.

\subsection{RIS aided NOMA wireless network}\label{section:2A}

As the combination of RISs and NOMA techniques is considered promising, a series of related research contributions have been proposed in the past years.  To combine the advantages of RISs and NOMA, authors of \cite{thirumavalavan2020ber} proposed a new RIS-aided downlink NOMA system to improve the reliability of the wireless network, and they derived the analytical expression of the bit error rate (BER) performance of RIS enhanced NOMA systems. The author of~\cite{yang2020secrecy} investigated the physical layer security of a RIS enhanced NOMA system. A NOMA based model of RIS-UAV communication was proposed in \cite{liu2020machine}, where the authors deployed RISs on the outer surface of the skyscraper to assist the wireless link of unmanned aerial vehicles (UAVs). UAVs' trajectories, passive beamforming of the RISs and power allocation are treated as optimization variables to minimize the energy cost of the UAVs. A partitioned RIS was employed in \cite{khaleel2020novel} to enhance the spectrum efficiency by improving the ergodic rate of all users, and the the physical resources distribution was optimized by three efficient search algorithms.
The authors of \cite{zhang2020joint} optimized user clustering, passive beamforming and power allocation for a downlike NOMA system with RISs by iteratively optimizing three sub-problems. An RISs enhanced NOMA cellular network with the joint transmission coordinated multipoint was proposed in~\cite{elhattab2020reconfigurable} to improve the data rate of edge users, while the network spectral efficiency was evaluated and validated through Monte-Carlo simulations. Meanwhile, in \cite{fu2019reconfigurable} and \cite{li2020joint}, joint optimizations of the base station beamforming and the passive beaming at the RIS were proposed with the aim of minimizing the total transmit power of the base station.

\subsection{DRL \& FL in wireless networks}\label{section:2B}

DRL has demonstrated commendable performance in various wireless network systems \cite{wang2020thirty}. By invoking a DRL approach, the authors of \cite{huang2020reconfigurable} investigated the joint design of transmit beamforming at the base station and the phase shifts at the RIS.
The author of \cite{yang2020intelligent} proposed a hill-climbing algorithm to optimize the power allocation at the base station and reflecting beamforming to achieve an anti-jamming communication.
Similar with \cite{liu2020machine}, an RISs assisted UAV communication system was invoked in \cite{wang2020joint} that a UAV and multiple RISs were paired to serve a number of ground users and two DRL approaches were adopted to maximize the overall weighted data rate and geographical fairness of by optimizing the UAV's trajectory and phase shifts of RISs. The authors of \cite{he2019joint} proposed a deep reinforcement learning approach attention-based neural network (ANN) to allocate resources for a multi-carrier NOMA system.

On the other hand, some researches on FL in wireless networks have been proposed \cite{9330566}. The author of \cite{elbir2020federated} proposed an FL approach to estimate channel for a RISs assisted massive Multi-input Multi-output (MIMO) system. To optimize the data rate of RISs aided networks, the authors of \cite{ma2020distributed} proposed an FL based beam reflection optimization algorithm to achieve high speed communication with the sparse CSI.
In addition, since the FL process needs to exchange data between agents via wireless networks, researches on how to use wireless communications to support federated learning is also challenging \cite{chen2019joint,yang2019energy,wang2020federated}. The authors of \cite{yang2019energy} formulated the joint learning and communication problem, and proposed an iterative algorithm to minimize the total energy consumption for an FL based system.

\section{System Model}\label{section:3}

In this section, we first describe assumptions and system model of the proposed mobile RISs enhanced wireless networks in \textit{subsection \ref{section:3A}}. The layout modeling method of the indoor environment and the propagation model are illustrated in \textit{subsection \ref{section:3B}} and \textit{subsection \ref{section:3C}}, respectively. The signal models of both OMA and NOMA scheme are illustrated in \textit{subsection \ref{section:3D}}. At last, the optimization problem is formulated in \textit{subsection \ref{section:3E}}.

\subsection{System Description and Assumption}\label{section:3A}

\begin{figure*}[t!] 
\centering 
\includegraphics[width=0.9\textwidth]{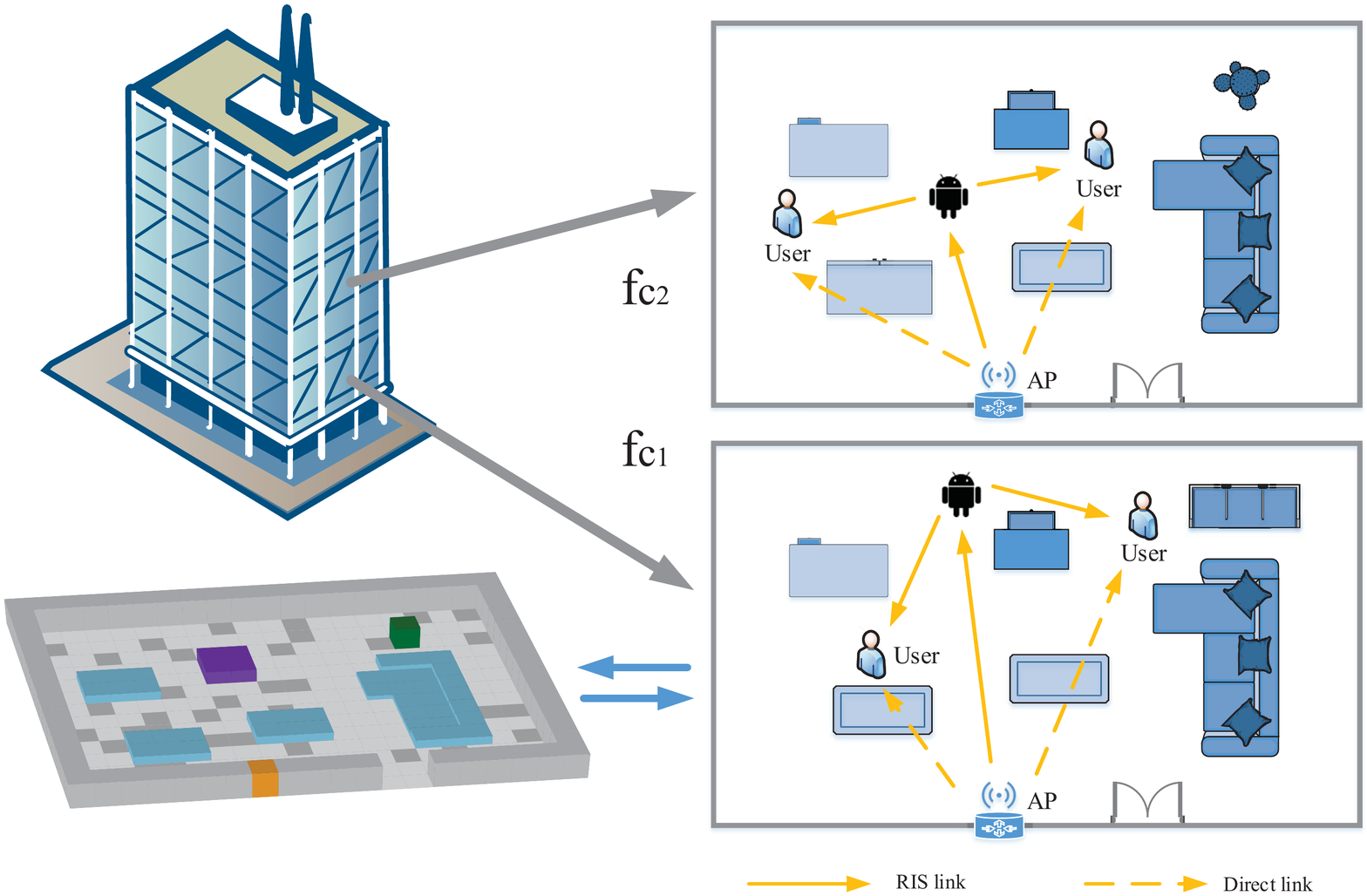} 
\caption{System model of NOMA enhanced mobile RIS} 
\label{Fig.main2} 
\end{figure*}

We consider an indoor downlink multiple-input and single-output (MISO) scenario where RISs are employed and each RIS is carried by a robot to enhance indoor propagation for a wireless access point (AP) to serve users in the room as illustrated in Fig.~\ref{Fig.main2}. We assume that the served building has multiple floors or rooms, we can denote each of them as a cell, and each cell is configured with an AP. In order to relieve the interference between each floor, we adopt a spectrum strategy similar to what is adapted in the cellular networks to diminish adjacent cell interference. The frequency band of the system is divided into at least two, and then adjacent floors can use different frequency bands. For example, if the frequency band is divided into two, odd-numbered floors can occupy the same frequency band, and even-numbered floors have to apply the other frequency band. Since the floors using the same frequency band are guaranteed to have a sufficient spatial distance and the signal is obstructed by ceilings, so that the interference between APs can be reduced to a negligible level.

We assume that each floor of the building has a similar architectural structure and layout, which is common in office buildings or flats. For each cell, the AP is equipped with $M$ antenna, while each user only has a single antenna. The RIS is armed with $N$ reflecting elements, which can provide concatenated LoS propagation for the transmitter and receivers by reflecting and reconfiguring the signals. We consider that there are multiple users in the room and they follow independent random movements \cite{cao2018secrecy}. Since users are constantly roaming, in order to maximize the channel gain, the robot carried RISs have to be deployed opportunely according to the real-time user distribution. The robot operates on the floor and the RIS is set at a fixed height on the robot, as a result the altitude of RIS is considered as a constant. In order to ensure safe operations, the robot cannot cross or collide with any obstacles, it also has to be guaranteed that the RIS will not collide with people.

\begin{remark}
The fixed-position RIS is likely to encounter blind spots when it is employed in indoor scenarios since furniture and room structures form a complex sheltered environment. Whether the RIS is mounted on the wall or ceiling, the LoS blind zone may be caused by girders, pillars, or chandeliers, and users in the blind zone can only get the NLoS channel. On the contrary, the RIS mounted on the robot can be deployed timely according to the user's location, which can improve the probability of LoS propagation for users.
\end{remark}

In this model, We denote the set of APs as $u \in \mathbb{U}=\{1,2,3...U\}$ and the set of users associated with AP $u$ as $k_u \in \mathbb{K}_u=\{1,2,3...K_u\}$. Users have to be associated with the AP on the same floor and the RISs employed is denoted as $r \in \mathbb{R}=\{1,2,3...R\}$. For a clear expression, we default the AP, RIS and the agent employed in the same cell have a corresponding order, for instance, if the AP order is $u=1$, the RIS working with $u$ is $r=1$. To express the concatenated propagation caused by RIS, we denote $\bm{h}_{u,r} \in \mathbb{C}^{M\times N}$ as the channel matrix of the link between AP and RIS and $\bm{h}_{r,k} \in \mathbb{C}^{N \times 1 }$ as the link of the $r$-th RIS to users. On the other hand, users can also receive the signal via the direct link (AP to user link). Thus, the channel between AP $u$ and user $k$ can be denoted as $\bm{h} \in \mathbb{C}^{M\times1}$.

In this paper, the passive beamforming of RISs is considered as one of the main optimization variables. Thus, we denote $\mathbf{\Theta}_r = \text{diag}[\beta_1e^{j\theta1}...\beta_ne^{j\theta_n}...\beta_Ne^{j\theta_N}]$ as the phase shift matrix of the RIS, where $\beta_n$ represents the amplitude of complex reflection
coefficient and $\theta_n \in [0,2\pi)$ represents the phase shift. On the contrary, since the main research scope of the paper is the joint optimization of the deployment and passive beamforming of RIS, the active beamforming at the AP side is solved by a conventional zero-forcing beamforming \cite{huang2019reconfigurable}.


\subsection{Interior Layout Modeling}\label{section:3B}

Prior to discussing indoor propagation and RISs' deployments, it is necessary to establish an interior layout model. With the assistance of the layout model, we would be able to determine whether there is LoS path between any two points in the indoor environment, which is one of the key knowledge for RISs to obtain significant channel gains.

In order to accurately represent the outline of the furniture, a number of fictitious bricks are engaged to construct the layout model instead of simple columns. For example, a digitized layout model for the office is shown in the lower left corner of Fig. \ref{Fig.main2}. Please note that theoretically this modeling method can describe any shape or object, but it will lead to a rise of computational complexity since each virtual brick has to be traversed to determine whether it occludes the LoS path.


\subsection{Propagation Model}\label{section:3C}

In order to simulate the indoor propagation, we do not invoke the statistical prorogation model since there is only a close range for indoor transmission distance (in metres) and a deterministic prorogation model is more conducive to precise planning the path of the carrier robot. Thus, we employ the aforementioned interior layout model and the indoor propagation model proposed by ITU recommendation ~\cite{ITUR} to obtain a deterministic indoor prorogation model.

We consider a propagation model including path loss and small-scale fading, which can be express as
\begin{align}\label{PL}
\mathcal{L}_{k_u}^ {u}(d) =  L_{k_u}^ {u}(d) - 10\log_{10}{h_{k_u}^ {u}},
\end{align}
where $h_{k_u}^ {u}$ denotes the Rician fading and $L_{k_u}^ {u}(d)$ represents the pass loss described in~\cite{ITUR,ITUR2}. With the aid of interior layout model and intersection detection~\cite{majercik2018ray}, we can calculate whether the link enjoys LoS. Then we can obtain deterministic pass loss
\begin{align}
L_{k_u}^ {u}(d)=
\begin{cases}
{L_\text{LoS}}(d), & \text{if LoS}, \\
{L_\text{NLoS}}(d,n), & \text{if NLoS}.
\end{cases}
\end{align}

For the NLoS link, the path loss can be calculated as
\begin{align}\label{PLNLOS}
{L_\text{NLoS}}(d,n) = L_0 + N\log_{10}{d} + L_f(n) ,
\end{align}
where variable $d$ represents the separation distance between the transmitter and the receiver and $n$ represents the the number of completely blocked obstacles, such as the walls or floors. $N$ denotes the distance power loss coefficient, as suggested in~\cite{ITUR}, we choose $N=25.5$ for the proposed office scenario. The parameter $f$ represents the carrier frequency in MHz. Please note that although we invoke discrepant frequency bands on adjacent floors, these center frequencies have to be adjacent to avoid the heterogeneity in transmission characteristics.

The term $L_0$ represents the basic transmission loss that can be calculated as
\begin{align}\label{L}
L_0 = 20\log_{10}{f} - 28,
\end{align}
and
\begin{align}
L_f(n)=15 + 4(n-1).
\end{align}

The path loss for the LoS link can be calculated as
\begin{align}\label{PLLOS}
{L_\text{LoS}}(d) = 16.9\log_{10}{d} - 27.2 + 20\log_{10}{f} .
\end{align}

\subsection{Signal Model}\label{section:3D}

\subsubsection{OMA Scheme}
In each cell, an FDMA scheme is adopted, and in order to further spectrum utilization, some users utilize the same frequency band. For users in the same frequency band, we apply zero-forcing beamforming to eliminate interference. The pre-coded transmitting signal from AP $u$ can be express as
\begin{align}\label{xt}
{x^u(t) = \sum\limits_{k_u = 1}^{K_u} {\sqrt {P_{k_u}^u(t)} } \bm{g}_{k_u}^{u}(t) s_{k_u}^{u}(t) },
\end{align}
where $s_{k_u}^{u}(t)$ represents the data symbol sequence from AP $u$ to user $k_u$ and $P_{k_u}^u(t)$ is the allocated power for user $k_u$. $\bm{g}_{k_u}^{u} \in \mathbb{C}^{M \times 1 }$ represents the active beamforming vector. Obviously these parameters are time-variant, so the time symbol $(t)$ is omitted in the subsequent equation to achieve a concise expression.

Thus, the received signal at user $k$ can be calculated as
\begin{align}\label{yt}
{y_{k_u} = (\bm{h}_{u,k_u} + \bm{h}_{r,k_u} \mathbf{\Theta}_r \bm{h}_{u,r})\sum\limits_{k_u = 1}^{K_u} {\sqrt {P_{k_u}^u} } \bm{g}_{k_u}^{u} s_{k_u}^{u} } + n_0,
\end{align}
where $n_0$ denotes the additive white Gaussian noise (AWGN) and for brevity of the express, in most cases path loss $L_{k_u}^u(t)$ in the rest of the paper is implicitly included in $\bm{h}_{u,k_u}$. As aforementioned, the active beamforming matrix $ \bm{g}_{j_u}^{u}$ is derived by a zero-forcing approach to mitigate the interferences. Thus, for a given user $k_u$ and interference user $j_u$ the pre-coding matrix can be calculated as
\begin{align}\label{ZF}
\begin{cases}
(\bm{h}_{u,k_u} + \bm{h}_{r,k_u} \mathbf{\Theta}_r \bm{h}_{u,r}) \bm{g}_{k_u}^{u} = 1, & \\
(\bm{h}_{u,j_u} + \bm{h}_{r,j_u} \mathbf{\Theta}_r \bm{h}_{u,r}) \bm{g}_{j_u}^{u} = 0, & j_u \neq k_u.
\end{cases}
\end{align}

We denote the ZF pre-coding matrix of AP $u$ as
\begin{align}
\bm{G_u} = [\bm{g_1^u}, ... \bm{g_{k_u}^u},... \bm{g_{K_u}^u}],
\end{align}
and if we denote $\bm{H}_{u,k_u} = [\bm{h}_{u,1},...,\bm{h}_{u,K_u}]$ and $\bm{H}_{r,k_u} = [\bm{h}_{r,1},...,\bm{h}_{r,K_u}]$ as a result, the direct channel and the concatenated channel can be regarded as an overall channel response as
\begin{align}
\bm{H_u} = \bm{H}_{u,k_u} + \bm{H}_{r,k_u}\mathbf{\Theta}_r \bm{H}_{u,r} .
\end{align}

Thus, the pre-coding matrix $\bm{G^u}$ can be calculated as the pseudo-inverse of overall channel response $\bm{H^u}$
\begin{align}
\bm{G_u} = \bm{H_u} {(\bm{H_u}^H \bm{H_u})}^{-1},
\end{align}

Therefore, based on (\ref{yt}) the signal-to-interference-plus-noise (SINR) for user $k$ can be calculated as
\begin{align}\label{SINROMA}
\gamma _{k_u} = \frac{\mid(\bm{h}_{u,k_u} + \bm{h}_{r,k_u} \mathbf{\Theta}_r \bm{h}_{u,r}) {\sqrt {P_{k_u}^u} } \bm{g}_{k_u}^{u} s_{k_u}^{u}\mid^2} {\mid(\bm{h}_{u,k_u} + \bm{h}_{r,k_u} \mathbf{\Theta}_r \bm{h}_{u,r})\sum\limits_{j_u \neq k_u} {\sqrt {P_{j_u}^u} } \bm{g}_{j_u}^{u} s_{j_u}^{u}\mid^2 + \sigma^2},
\end{align}
where $\sigma^2$ is the average power of the AWGN \footnote{If the multiple access approach is assumed to be ideally orthogonal, the inter-user interference can be considered as zero.}. Consequently, the data rate of user $k_u$ at time $t$ can be calculated as
\begin{align}
\mathcal{R}_{k_u}^u = B_{k_u}\log 2\left( {1 + \gamma _{k_u}} \right).
\end{align}

\subsubsection{NOMA Scheme}

Contrary to the OMA scheme, the NOMA technique allows multiple users to form a cluster and utilize the same frequency band simultaneously. Hence, for each user cluster $v \in \mathbb{V}= (1,2...V)$, and we denotes the users in cluster $v$ as $k_v$. We also assume that the maximum callable power of each cluster is the same, and the transmitted signal can be expressed as
\begin{align}\label{xnu}
{x^v = \sum\limits_{k_v = 1}^{K_v} {\sqrt {P_{k_v}^u(t)} } s_{k_v}^u(t) },
\end{align}
and then the transmitting signal of AP can be expressed as
\begin{align}\label{xuNOMA}
{x^u = \sum\limits_{v = 1}^{V} \bm{g}_{v}^{u} \sum\limits_{k_v = 1}^{K_v} {\sqrt {P_{k_v}^u(t)} } s_{k_v}^u(t) } ,
\end{align}
where $\bm{g}_{v}^{u}$ represents the ZF pre-coding matrix.
Therefore, the received signal of user $k$ in the NOMA cluster $v$ served by AP $u$ can be expressed as
\begin{align}
y_{k_u} = (\bm{h}_{u,k_v} + \bm{h}_{r,k_v} \mathbf{\Theta}_r \bm{h}_{u,r}) \bm{g}_{v}^{u} x^u_{k_v} + I^v_{k_v} + I^u_{k_v} + n_0,
\end{align}
where $(\bm{h}_{u,k_v} + \bm{h}_{r,k_v} \mathbf{\Theta}_r \bm{h}_{u,r}) x^u_{k_v}$ is the desired signal of user $k_v$. $I^v_{k_v} $ denotes the intra-cluster interference and $I^u_{k_v} $ denotes the inter-cluster interference received by user $k_v$.

The inter-cluster interference can be calculated as
\begin{align}
I^u_{k_v} = \sum\limits_{\mathfrak{v} = 1,\mathfrak{v} \ne v}^V {(\bm{h}_{u,k_v} + \bm{h}_{r,k_v} \mathbf{\Theta}_r \bm{h}_{u,r}) \bm{g}_{\mathfrak{v}}^{u} x^\mathfrak{v}}.
\end{align}

In order to obtain comparable results, the same ZF beamforming is also invoked at the NOMA AP. Similar with the OMA case, the pre-coding for NOMA can be expressed as
\begin{align}\label{ZFNOMA}
\begin{cases}
(\bm{h}_{u,k_v} + \bm{h}_{r,k_v} \mathbf{\Theta}_r \bm{h}_{u,r}) \bm{g}_{k_v}^{u} = 1, & \\
(\bm{h}_{u,j_\mathfrak{v}} + \bm{h}_{r,j_\mathfrak{v}} \mathbf{\Theta}_r \bm{h}_{u,r}) \bm{g}_{j_\mathfrak{v}}^{u} = 0, & j_\mathfrak{v} \neq k_v, v \neq \mathfrak{v}.
\end{cases}
\end{align}

The derivation process of the pre coding matrix for NOMA is the same as the OMA scheme, thus we can also obtain it as
\begin{align}
\bm{G_v^u} = \bm{H_{u,v}} {(\bm{H_{u,v}}^H \bm{H_{u,v}})}^{-1}.
\end{align}

It can be observed in (\ref{xuNOMA}) that instead of design beamforming for each individual user in the OMA scheme, a beam in the NOMA system is designed for a NOMA cluster.
Since users in the same NOMA cluster are also likely to have different channel responses, ZF beamforming cannot eliminate inter-cluster interference for all users in a cluster. For example, assuming users $j_v$ and $i_v$ are in the same cluster $v$ with channel $(\bm{h}_{u,j_v} + \bm{h}_{r,j_{v}} \mathbf{\Theta}_r \bm{h}_{u,r})  \neq (\bm{h}_{u,i_v} + \bm{h}_{r,i_{v}} \mathbf{\Theta}_r \bm{h}_{u,r})  $. According to (\ref{ZFNOMA}) we have $(\bm{h}_{u,j_v} + \bm{h}_{r,j_{v}} \mathbf{\Theta}_r \bm{h}_{u,r}) \bm{g}_{k_{\mathfrak{v}}}^{u} = 0$ and it is easy to figure out $(\bm{h}_{u,i_v} + \bm{h}_{r,i_{v}} \mathbf{\Theta}_r \bm{h}_{u,r}) \bm{g}_{k_{\mathfrak{v}}}^{u} \neq 0$, which suggests the inter-cluster interference cannot be remove completely at user $j_v$. We select the user with the highest channel power gain which we call it the strangest user in each cluster as the basis for the beamforming in order to keep a correct decoding order for SIC. Therefore, the inter-cluster can be dislodged at the strongest user in each cluster but weaker users still have to suffer.

On the other hand, the a portion of intra-cluster interference can be eliminated by SIC and the intra-cluster interference for each user can be calculated with a given decoding order. Since users and RISs keep moving, the channel quality will be fickle, so a dynamic decoding order has to be determined in each time slot. For the convenience of presentation, we assume that users in NOMA cluster $v$ have a consistent numbering order with channel quality at time $t$, where user $K$ is the strongest user. Consider user $j_v$ and $k_v$ at time $t$ have relationship that
\begin{align}
\frac{\mid(\bm{h}_{u,j_v} + \bm{h}_{r,j_{v}} \mathbf{\Theta}_r \bm{h}_{u,r})\mid}{\sum\limits_{\mathfrak{v} = 1,\mathfrak{v} \ne v}^V {\mid(\bm{h}_{u,j_v} + \bm{h}_{r,j_v} \mathbf{\Theta}_r \bm{h}_{u,r}) \bm{g}_{\mathfrak{v}}^{u} x^\mathfrak{v}}\mid L_{u,k_v}} >
\frac{\mid(\bm{h}_{u,k_v} + \bm{h}_{r,k_{v}} \mathbf{\Theta}_r \bm{h}_{u,r})\mid}{\sum\limits_{\mathfrak{v} = 1,\mathfrak{v} \ne v}^V {\mid(\bm{h}_{u,k_v} + \bm{h}_{r,k_v} \mathbf{\Theta}_r \bm{h}_{u,r}) \bm{g}_{\mathfrak{v}}^{u} x^\mathfrak{v}}\mid L_{u,j_v}},
\end{align}
where $L_{u,k_v}$ represents the path loss in linear. After that, user $j_v$ can adopt SIC to remove the signal for user $k_v$ in prior of decoding the signal for itself \cite{cui.signal}. Thus, the decoding order can be denote as $k_v<j_v$. Therefore, generalize the above theory to multi-user clusters, the decoding order for cluster $v$ at time $t$ can be expressed as $\{1,2...k...K\}$, where user $k_v$ is the k-th user to practice SIC decoding in this cluster. As a result, the intra-cluster interference at user $k$ can be calculated as
\begin{align}\label{inuk}
I^v_{k_v} = \sum\limits_{j_v = k_v+1}^{K_v} {(\bm{h}_{u,k_v} + \bm{h}_{r,k_v} \mathbf{\Theta}_r \bm{h}_{u,r}) \bm{g}_{K_v}^{u} x^v_{j_v}}.
\end{align}

Therefore, the received SINR for user $k_v$ can be calculated as
\begin{align}\label{SINROMA}
\gamma^u _{k_v} = \frac{\mid(\bm{h}_{u,k_v} + \bm{h}_{r,k_v} \mathbf{\Theta}_r \bm{h}_{u,r}) {\sqrt {P_{k_v}^u} } \bm{g}_{K_v}^{u} s_{k_v}^{u}\mid^2} {\mid(\bm{h}_{u,k_v} + \bm{h}_{r,k_v} \mathbf{\Theta}_r \bm{h}_{u,r})\sum\limits_{j_v = k_v+1}^{K_v} {\sqrt {P_{j_v}^u} } \bm{g}_{K_v}^{u} s_{j_v}^{u}\mid^2 +  \mid  \sum\limits_{\mathfrak{v} = 1,\mathfrak{v} \ne v}^V {(\bm{h}_{u,k_u} + \bm{h}_{r,k_u} \mathbf{\Theta}_r \bm{h}_{u,r}) \bm{g}_{\mathfrak{v}}^{u} x^\mathfrak{v}} \mid^2+ \sigma^2}.
\end{align}

At last, the data rate of user $k_v$ served by AP $u$ can be calculated as
\begin{align}
\mathcal{R}_{k_v}^u = B^u_{k_v}\log 2\left( {1 + \gamma^u _{k_v}} \right).
\end{align}

\subsection{Problem Formulation}\label{section:3E}

We aim to maximize the sum data rate of users by jointly optimizing robot-mounted RISs' deployment $\mathcal{D}_r = \{ D_r(1) , D_r(2), ... D_r(t) ... \}, r \in \mathbb{R},$ and the phase shift for all reflecting element $\bm{\Theta}_r = \{\bm{\Theta}_r(1),\bm{\Theta}_r(2),...\bm{\Theta}_r(t)...\}, r \in \mathbb{R}$ of the mobile RIS, where $D_r(t)=[x_r(t),y_r(t),z_r(t)]$ represents the position of mobile RIS $r$ at time $t$.
Meanwhile, since APs need to collaborate with RISs, the corresponding power allocation policy $\mathcal{P}_r= \{P^u(1),P^u(2),...P^u(t)\}, u \in \mathbb{U}$.
Thus, the optimization problem can be formulated as
\begin{subequations}
\begin{align}\label{OPP}
&\max_{\mathcal{D}_r,\mathcal{P}_r,\bm{\Theta}_r} \sum_{k_u=1}^{K_u} \sum_{u=1}^{U} \mathcal R_{k_u},\\
\textrm{s.t.} \ \
& {x_{\min }} \le x_r(t) \le {x_{\max }},\forall r,\forall t,\notag\\
& {y_{\min }} \le y_r(t) \le {y_{\max }},\forall r,\forall t \label{OPPD},\\
& \sum\limits_{k_v \in {\mathbb{K}_v }} { {P^u_{k_v}(t) \le {P}^u_{v\text{max}}} }, \forall t,\forall v,\forall u,\label{OPPE}\\
& k^u_v (t) < j^u_v(t), \forall (k,j),\forall t,\forall k,\forall u,\label{OPPB}\\
& \mathcal{R}(t) \geq \mathcal{R}_{\text{QoS}},\forall t,\forall k,\forall u,\label{OPPG}
\end{align}
\end{subequations}
where constraint \eqref{OPPD} ensures that the mobile RISs have to be deployed in the appointed room, since once a mobile RIS is moved to other areas, it may cause unexpected interference especially when multiple RISs are deployed in the same room. Constraint \eqref{OPPE} is a power constraint that the total power allocated to users in a cluster cannot exceed the maximum power that the cluster is authorized to invoke while in the OMA scheme a signal user can be regarded as a cluster.  Constraint \eqref{OPPB} is introduced to ensure that the user ordering and decoding order can be performed correctly in each NOMA cluster. Finally, taking into account the fairness of users, constraint \eqref{OPPG} represents the data rate of each user at any time $t$ is guaranteed to meet the minimum rate of QoS requirement. As mentioned above, the predicament of the optimization is that the formulated problem is dynamic, non-convex \cite{na2020join} and the obstructive environment is non-functional. It is worth mentioning that the phase shift optimization in the NOMA scenario not only provides channel enhancement for users, but the channel modification has to be NOMA-friendly as well to take care of user fairness. Therefore, a DRL algorithm is invoked to solve the formulated problem.



\section{Federated Learning Model}\label{section:4}

%
%
%
%
%
In \textit{subsection \ref{section:4A}}, we elaborate on the role and superiority of invoking FL to coordinate multiple agents and prove that there is a theoretical gain in FL-DRL framework. In \textit{subsection \ref{section:4B}}, we propose an FL model with local training to serve multiple cell networks.
\subsection{The Concept of FL}\label{section:4A}

Federated learning is competent to be invoked for optimizing the proposed communication model. As aforementioned, the proposed indoor network composed of APs has cellular characteristics to extend, and independent agents served in each cell have great common functions and attributes. For example, the pursuit of service quality, the equipment of RISs and the propagation characteristics of signals in each cell are equal, which constitutes the cornerstone of adopting the FL framework.

The FL framework has a number of common advantages for all ML algorithms, for example, it can save more hardware resources, improve training speed, and with the protection of user privacy \cite{qin2020federated}. Moreover, in addition to these frequently mentioned advantages, DRL algorithms are specifically suitable to be applied in the FL framework. The learning process of RL comes from continuously interacting with the environment and exploring different states and actions. However, the exploration of the environment is not likely to exhaust all states though the action policy contains random actions or noise \cite{lin2020deep,zhang2019discretionary}, which leads to the global optimum may being buried in the quagmire.  In particular, the proposed communication scenarios and indoor layouts have high complexity, impelling the efficient and sufficient exploration to be a problem. Therefore, the training effect of DRL is determined by whether the agent has sufficient exploration and experience, fortunately, the participation of FL is helpful to reveal more different states since multi-agents are investigating the environment, which allows the environment to be explored more sufficiently.

\begin{remark} \label{rmark.FL}
When the environments explored by the DRL agents have similarities and the state transitions have not been traversed by agents, the FL framework can provide potential gains than independent training scheme since it is likely to obtain more sufficient environmental knowledge.
\end{remark}
Proof of \textbf{Remark \ref{rmark.FL}}:

Assuming a Markov process has $\mathbf{S}$ states, each state $S \in \mathbf{S}$ has action space $\mathbf{A}^S_N$ and the corresponding reward set $\mathbf{R}^{(S,A)}_N$, denoting the explored action space of FL agents as $\mathbf{A}^S_F \subseteq\mathbf{A}^S_N$,$\mathbf{R}^{(S,A)}_F \subseteq \mathbf{R}^{(S,A)}_N$ and explored action space of the independent agent as $\mathbf{A}^S_I\subseteq\mathbf{A}^S_N$,$\mathbf{R}^{(S,A)}_I \subseteq \mathbf{R}^{(S,A)}_N$. Assuming that a repetitive tolerant random action policy is adopted during the exploration process, we can get $E[ \mid\mathbf{A}^S_F\mid] \geq E[ \mid\mathbf{A}^S_I\mid]$ and also for the reward set $E[ \mid\mathbf{R}^{(S,A)}_F\mid] \geq E[ \mid\mathbf{R}^{(S,A)}_I\mid]$. For the reward sets, the maximum known reward $\text{max}(\mathbf{R}^{(S,A)}_F,\mathbf{R}^{(S,A)}_I)$ always exists, though there may have $\text{max}(\mathbf{R}^{(S,A)}_F,\mathbf{R}^{(S,A)}_I)< \text{max}(\mathbf{R}^{(S,A)}_N)$.
Then the probability that the known maximum reward is found by FL agents and independent agents can be calculated as
\begin{align}
P[\text{max}(\mathbf{R}^{(S,A)}_F,\mathbf{R}^{(S,A)}_I)\in \mathbf{R}^{(S,A)}_I] = 1-(1-\frac{1}{\mid \mathbf{R}^{(S,A)}_N\mid_{\mathbf{R}^{(S,A)}_N\leq\text{max}(\mathbf{R}^{(S,A)}_F,\mathbf{R}^{(S,A)}_I)}})^{E[ \mid\mathbf{A}^S_I\mid]},
\end{align}
\begin{align}
P[\text{max}(\mathbf{R}^{(S,A)}_F,\mathbf{R}^{(S,A)}_I)\in \mathbf{R}^{(S,A)}_F] = 1-(1-\frac{1}{\mid \mathbf{R}^{(S,A)}_N\mid_{\mathbf{R}^{(S,A)}_N\leq\text{max}(\mathbf{R}^{(S,A)}_F,\mathbf{R}^{(S,A)}_I)}})^{E[ \mid\mathbf{A}^S_F]\mid}.
\end{align}

Since $E[ \mid\mathbf{A}^S_F\mid] \geq E[ \mid\mathbf{A}^S_I\mid]$, then
\begin{align}
P[\text{max}(\mathbf{R}^{(S,A)}_F,\mathbf{R}^{(S,A)}_I)\in \mathbf{R}^{(S,A)}_I] \leq P[\text{max}(\mathbf{R}^{(S,A)}_F,\mathbf{R}^{(S,A)}_I)\in \mathbf{R}^{(S,A)}_F], \forall S \in \mathbf{S},
\end{align}
and it can be obtained that
\begin{align}\label{E_FL}
E[\text{max}(\mathbf{R}^{(S,A)}_I)] \leq E[\text{max}(\mathbf{R}^{(S,A)}_F)], \forall S \in \mathbf{S}.
\end{align}

It is worth to point out that $E[\text{max}(\mathbf{R}^{(S,A)}_I)] = E[\text{max}(\mathbf{R}^{(S,A)}_F)]$ if $\mathbf{A}^S_F=\mathbf{A}^S_I=\mathbf{A}^S_N$, which reveals that when the action space is traversed, FL will no longer provide gain. Then, the cumulative reward which described by value function
\begin{align}\label{Q_func}
Q(S,A)
= R(S,A) + \beta \sum_{S'\in\mathbf{S}}{\max_{A'}Q(S',A')}.
\end{align}

According to (\ref{E_FL}) and (\ref{Q_func}), we can obtain
\begin{align}
E[\text{max}(Q_I(S,A))] \leq E[Q_F(S,A))].
\end{align}

For DQL algorithms, assuming the neuronal network fitting correctly, we have $Q^*\rightarrow Q$, then
\begin{align}
E[\text{max}(Q^*_I(S,A))] \leq E[Q^*_F(S,A))].
\end{align}

\textbf{Remark \ref{rmark.FL}} is proven.

Therefore, the FL scheme has more expected gain than independent agents until all state
transitions have been traversed by agents.
Furthermore, the global model can also greatly enriches the experience diversity since each agent has different initialization and pseudorandom. In summary, by establishing a global model and exchanging neural network parameters, agents located on different floors or cells can learn from each other's experiences. The introduction of FL can improve the training efficiency and effect of DRL algorithms and the gain is also revealed by the simulation results in Section \ref{section:6}.

\subsection{FL Model for DRL}\label{section:4B}

\begin{figure}[t!]
\centering
\includegraphics[width=0.8\columnwidth]{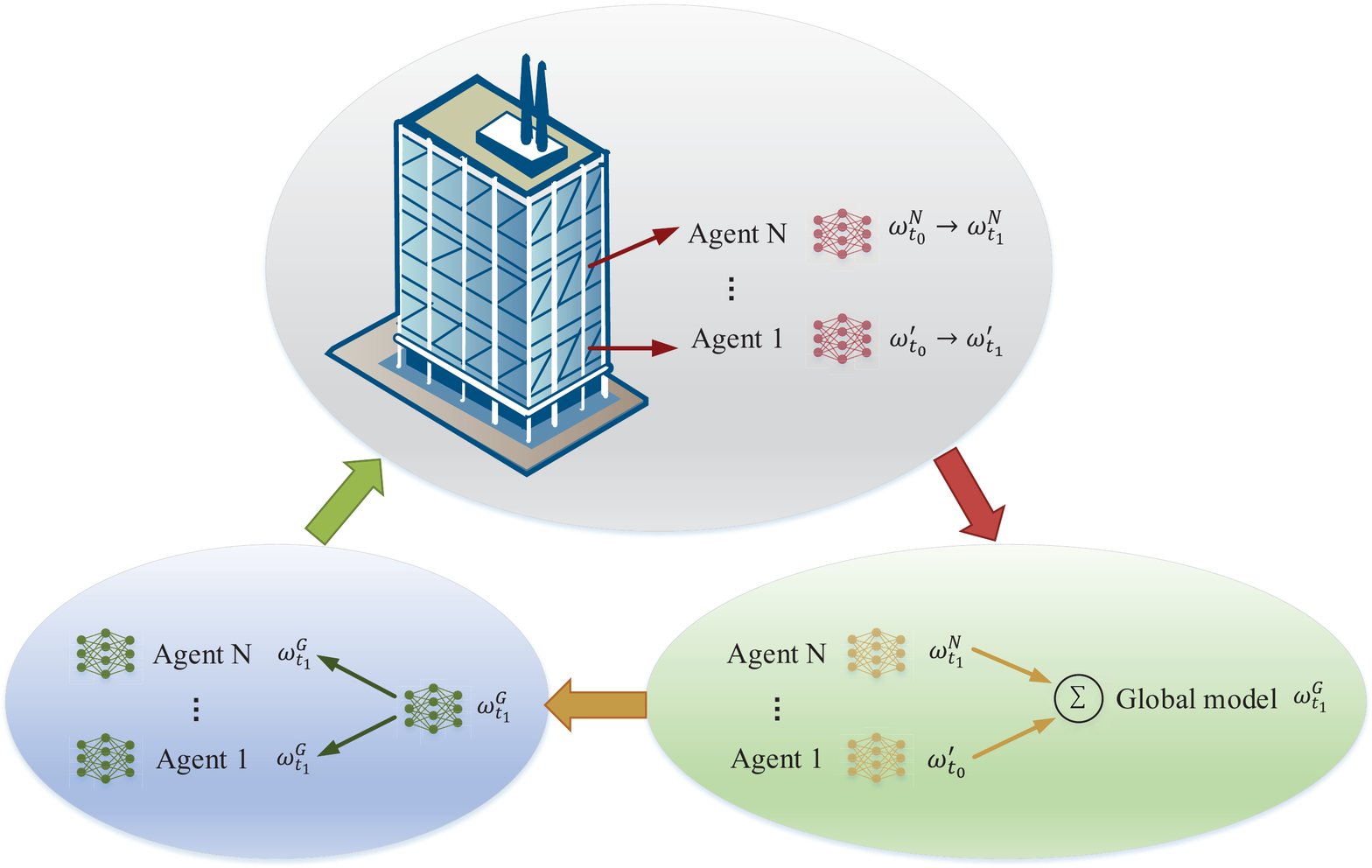}
\caption{Federated learning enhanced indoor mobile RIS network}\label{Fig.FL}
\end{figure}

Our FL framework adopts decentralized training and uses federated averaging to generate a globe model. The operation process can be divided into three parts: local training, updating global model, and downloading global model, which is illustrated in Fig. \ref{Fig.FL}.
\begin{itemize}
\item Local training: Each local agent set up their local model $\omega_t^u$ and uses its own computing resources to train the local model, where $t$ represents the time and $u$ represents the agent number. Local neural network models have random initialization to increase the diversity of exploration at early training.
\item Global model update: After a period of training interval $F_G$, the parameters of the global model $\omega_t^G$ can be upgraded by averaging the parameters of each local model, which can be express as
    \begin{align}
    \omega_t^G = \frac{1}{U} \sum_{u=0}^{U}{\omega_t^u}.
    \end{align}
\item Local model update: After the global model is updated, each agent downloads the global model and then updates the local model according to the global model.
    \begin{align}
    \omega_t^u = \omega_t^G, \forall u.
    \end{align}
    After the updating is complete, the new model can be used for the next round of local training.
\end{itemize}

\section{FL-DDPG executed optimization for Mobile RISs}\label{section:5}
With the aforementioned FL framework, this section details the FL enhanced DDPG algorithm to optimize the deployment, phase shifts of RISs, and the power allocation for users. The algorithm training and decision flow is explained in \textit{subsection \ref{section:5A}}. As a DRL approach, the specific state space and action space design for mobile RIS scenario is presented in \textit{subsection  \ref{section:5B}}, and the adaptive neural network structure is introduced in \textit{subsection  \ref{section:5C}}. Furthermore, \textit{subsection \ref{section:5D}} analyses the convergence and complexity of the FL-DDPG algorithm.

\subsection{FL-DDPG Algorithm and Training}\label{section:5A}
\begin{figure}[htbp]
\centering
\includegraphics[width=1\columnwidth]{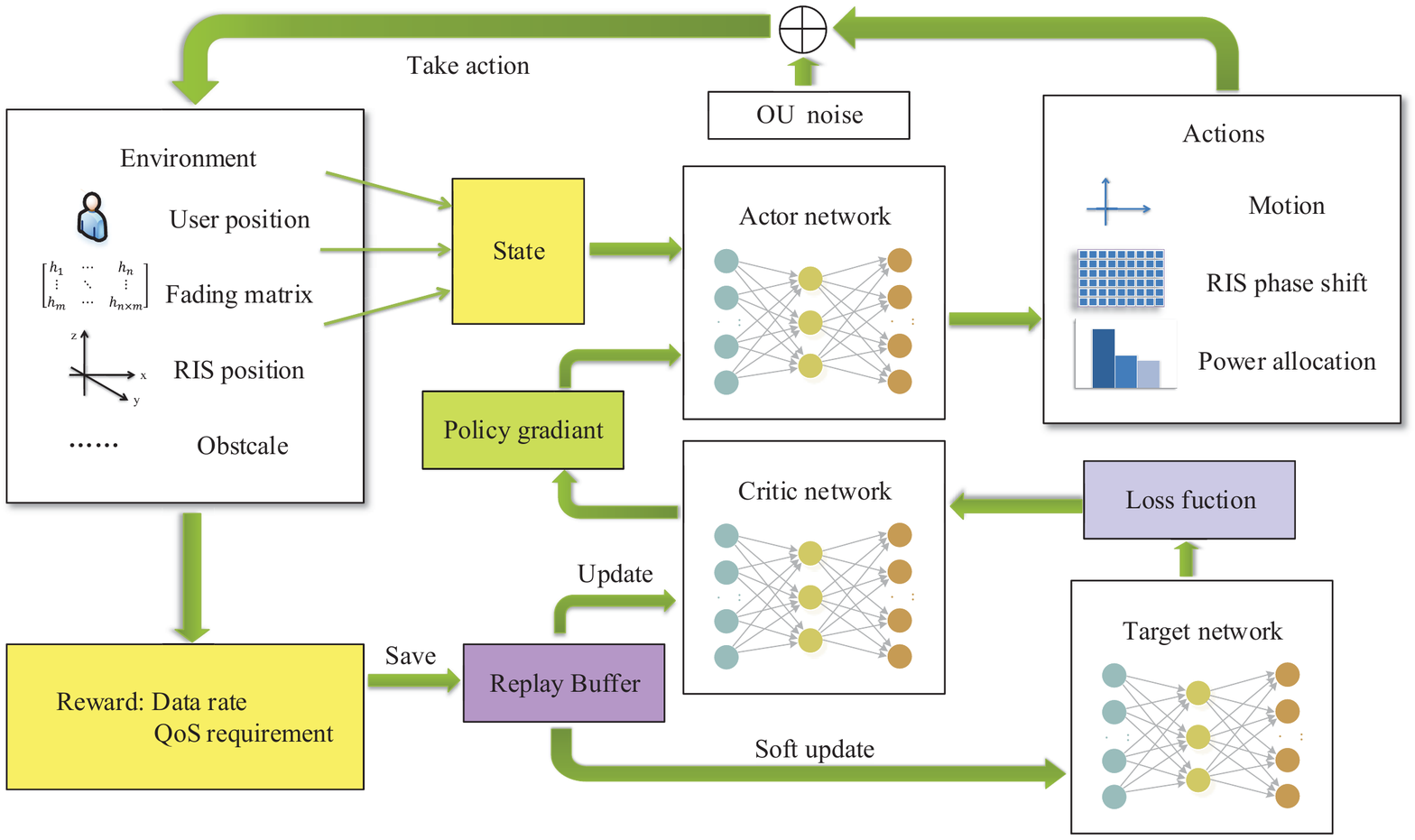}
\caption{Flow diagram of the local training in the FL-DDPG algorithm}\label{Fig.FLDDPG}
\end{figure}

We propose an FL-DDPG algorithm to jointly optimize deployments, phase shifts of mobile RIS and the corresponding power allocation policy for users in each cell. Additionally, we implement several improvements on the original DDPG algorithm \cite{lillicrap2015continuous}, such as decaying Ornstein Uhlenbeck (OU) noise and adaptive neural network structure to adapt the algorithm into the proposed communication scenarios. We assume that each local agent is deployed within the AP and it can control the actions of the RIS and the carrier robot via the control channel. Due to the actor-critic structure, four neural networks are used in the DDPG agent, namely the actor network $Q$, the critic network $\mu$, the actor target network $Q'$ and the critic target network $\mu'$. Once observing the environment state $S_t$, the actor network calculates the action $A_t$ and then it will be executed. After the action is executed, the state will be changed to $S_{t+1}$, and the reward $R_t$ will be calculated according to the data rate $\mathcal{R}_t$ and QoS requirement threshold. The detailed update flow of a single DDPG agent is presented in Fig. \ref{Fig.FLDDPG}.

In order to train the agent efficiently, we adopted decaying OU noise in the training process
\begin{align}\label{Action}
A_t = \mu(S_t|\omega_t^\mu)+N(0,\xi_t), \xi_t = \xi_0 \rightarrow 0, \xi_0 \in [1,0),
\end{align}
where $\omega_t^\mu$ represents the parameters of neural network $\mu$ and $\xi_t $ denotes the scale of the OU noise. The OU action noise can drive the agent to explore further diversely compare to the Gaussian noise \cite{colas2018gep}, and decreasing noise can improve exploration efficiency without loss of convergence. On the other hand, memory replay technology is adopted in our model. The agent record and store the transition $(S_t,A_t,R_t,S_{t+1})$ for each step into a replay memory buffer and randomly sample experiences at each step and train neural networks according to the samples. For single sample at each step, the actor network can be updated according to the policy gradient. Assuming the  minibatch has $e$ transition samples, the policy gradient can be calculated as
\begin{align}\label{Actor}
\nabla_{\omega^\mu}J = \frac{1}{e}\sum_{e}\nabla_A Q(S_{t=e},A_{t=e}|\omega^Q)\nabla_\omega^\mu \mu(S_{t=e}|\omega^\mu).
\end{align}

The critic network is in charge of evaluating the action value (Q-value) of the action taken actions taken in a certain state, which is similar as the Q-learning and deep Q-network (DQN) algorithms. A Q-value with a concern of long-term reward is defined by the Bellman equation
\begin{align}\label{Bellman}
Q(S_t,A_t) = R_t(S_t,A_t) + \beta \max Q(S_{t+1},A_{t+1}).
\end{align}

In order to accurately estimate Q-value, the critic network is updated by minimizing the loss function
\begin{align}\label{Loss}
L_e = \frac{1}{e}\sum_{e}(y_{t=e} - Q(S_{t=e},A_{t=e}|\omega^Q))^2,
\end{align}
where
\begin{align}\label{Y}
y_t = R_t(S_t,A_t) + \beta  Q'(S_{t+1},\mu'(S_{t+1}|\omega^{\mu'})|\omega^{Q'}).
\end{align}

\begin{algorithm}
\caption{FL-DDPG algorithm for the sum rate optimization}
\label{DTDPG}
\begin{algorithmic}[1]

        \FOR{each cell $u \in \mathbb{U}$}
        \STATE Initialize the environment and determine the neural network specifications based on the number of RIS elements
        \STATE Initialize the actor network $\omega^Q_u$, critic network $\omega^{\mu}_u$, target actor network $\omega^{Q'}_u$, target critic network $\omega^{\mu'}_u$ with random parameters
        \FOR{each episode $\mathcal{E}$}
        \IF {$\mathcal{E}$ \% $F_G$ =0}
            \STATE Update global model $\omega_G^{Q,Q',\mu,\mu'} = \frac{1}{U} \sum_{u=0}^{U}{\omega_u^{Q,Q',\mu,\mu'}}$
            \STATE Update local models $\omega_u^{Q,Q',\mu,\mu'} = \omega_G^{Q,Q',\mu,\mu'}.$
        \ENDIF

        \STATE Reset the environment and initial state
        \FOR{each step in $ t_0 \leq t \leq t_\text{max}$}
        \STATE Observe $S_t$ according to the radio map
        \STATE Choose $A$ according to action policy and $Q(S,\omega^Q)$
        \STATE IRs take action $A$, observe $R_t$ and $S_{t+1}$
        \STATE Record $e \{S_t,A,R,S_{t+1}\}$
        \STATE Random sample a batch of transection $e$ from memory buffer

        \STATE Calculate target according to \eqref{Y}
        \STATE Train critic network $\mu(S,\omega^\mu)$ with a gradient descent step \eqref{Loss}
        \STATE Train actor network ${Q}(S,\omega^{Q})$ with \eqref{Actor}
        \STATE Update the target networks $\omega^{Q'}\leftarrow (1-\tau)\omega^{Q'} + \tau \omega^Q$, $\omega^\mathcal{\mu'}\leftarrow (1-\tau)\omega^\mathcal{\mu'} + \tau\omega^\mathcal{\mu}$
        \STATE $S_t \leftarrow S_{t+1}$
        \ENDFOR
        \STATE Each agent save the network models $\omega^{Q}_u$, $\omega^\mu_u$, $\omega^{Q'}_u$, $\omega^{\mu'}_u$
        \ENDFOR
		\ENDFOR

\end{algorithmic}
\end{algorithm}

\subsection{State, Action and Reward Function}\label{section:5B}

The DDPG algorithm supports continuous state and action space. Therefore, regardless of movement, phase shifts and power allocation are designed to be continuous to obtain the accurate action and we design the following state, action space and reward function.

\subsubsection{State Space}

For a single agent, the state space $S_t$ contains four components, the RIS location $D_r(t)$ in time slot $t$,  user location $D_{k_u}(t), k_u \in \mathbb{K}_u$, pass loss for each user $L_{k_u}^ {u}(t)$ and fading matrixes $\bm{h}_{u,k},\bm{h}_{u,r}$ and $\bm{h}_{r,k}$. Thus, the state for time slot $t$ can be noted as
\begin{align}\label{S}
S_t = \{D_{r}(t),D_{k_u}(t),L_{k_u}^ {u}(t), \text{real}\{\bm{h}_{u,k},\bm{h}_{u,r},\bm{h}_{r,k}\}, \text{imag}\{\bm{h}_{u,k},\bm{h}_{u,r},\bm{h}_{r,k}\}\}, k_u \in \mathbb{K}_u.
\end{align}

Since the elements in fading matrixes are complex numbers, the real and imaginary parts of each element can be split and input to different nodes. These selected parameters are necessary, while the deployment plan requires location information, and the optimization of power allocation and phase shifts is based on CSI. In addition, since the state is composed of unrelated variable categories, their values may have a colossal gap, and therefore proper scaling is necessary to avoid some values being ignored.

\subsubsection{Action Space}

The composition of the action space completely corresponds to the three optimization parameters, including motion, phase shift and power allocation.

\begin{itemize}
\item Deployments: For the deployment, the agent does not calculate the optimal position but choosing the next move $\Delta D_{r}(t)$ for the robot at each time slot $t$. The proposed approach allows the agent to find the optimal movement at each moment, with a consideration of long-term reward. However, the method of directly finding an optimal position will cause the moving path of mobile RIS may not be optimal.
\item Phase shifts: The agent calculates the optimal $\bm{\Theta}_r(t)$ at the current moment for each element express them in a radian system. The time for rotating the angle of reflecting elements is neglected.
\item Power allocation policy: The agent allocate power $P_{k_u}^ {u}(t)$ to each associated user $k_u$ at each time slot, where the allocated power meets $P_{k_u}^ {u}(t) < P_{\text{max}k_u}^ {u}(t)$. For the OMA scenario, $P_{\text{max}k_u}^ {u}(t)= P_{\text{max}}^{u}/K_u$ but in NOMA cases users can have their own power upper bound while $\sum\limits_{k_u}P_{\text{max}k_u}^ {u}(t) \leq P_{\text{max}}^{u}/K_u$.

\end{itemize}
In summary, the action space can be noted as
\begin{align}\label{S}
A_t = \{\Delta D_{r}(t),\bm{\Theta},P_{k_u}^ {u}(t)\},k_u \in \mathbb{K}_u.
\end{align}

\subsubsection{Reward Function}

For each cell, in order to maximize the data rate of the system, the reward is set to be proportional to the sum rate of all users. As mentioned in (\ref{OPPG}), in order to meet the user fairness constraint, once the data rate of any user does not meet the QoS requirements, a penalty has to be imposed. The agent will receive a discounted reward as in (\ref{R}), where $\lambda$ is the reduction factor

\begin{equation}\label{R}
R_t=
\begin{cases}
\mathcal{R}^u(t), &  \text{QoS requirement satisfied},\\
\frac{\mathcal{R}^u(t)}{\lambda}, &  \text{QoS requirement not satisfied}.
\end{cases}
\end{equation}

\subsection{Neural Network Structure for DT-DPG algorithm}\label{section:5C}
\begin{figure}[t!]
\centering
\includegraphics[width=0.6\columnwidth]{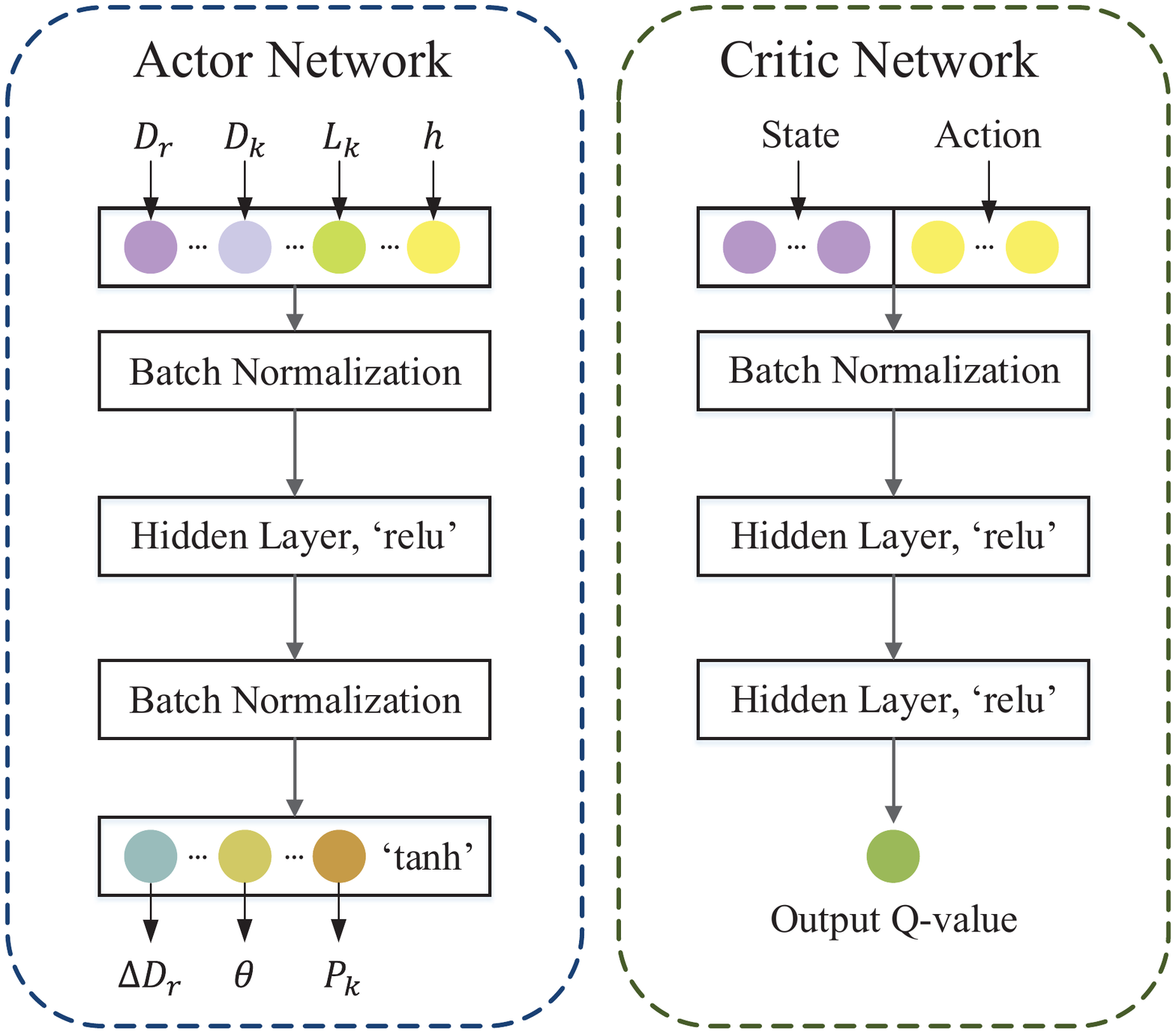}
\caption{Neural network structure of the proposed FL-DDPG algorithm} \label{Fig.NN}
\end{figure}

The structures of the actor network and the critic network is presented in Fig. \ref{Fig.NN}. Two batch normalization (BN) layers and an activation layer with relu function are employed in the actor network. The first BN layer is in charge of normalizing input data and the second BN layer ensures a valid input range for the tanh layer. Since all elements of fading matrices $\bm{h}_{u,k},\bm{h}_{u,r}$ and $\bm{h}_{r,k}$ need to be input to the actor network as the basis for the phase shift optimization, then the size of the input dimension have to be adaptive and determined by the number of users and the number of elements in RIS. The size of the hidden layer should also be adjusted accordingly to the communication system to achieve a proper fitting effect. The empirical number of the activation layer nodes is $\omega_{\text{relu}} = 4MN$, where the position input is not counted since it adds a negligible input dimension. A similar structure is adopted in the critic network. Since the critic network only needs to output a Q-value, its hidden layers can have a minor size, although the critic network has a larger input dimension.

\subsection{Convergence and Complexity Analysis}\label{section:5D}

The convergence of the basic Q-learning has been proved in a series of literatures, such as \cite{melo2001convergence}. However, due to the introduction of neural networks, the convergence of the DDPG algorithm is no longer guaranteed \cite{lillicrap2015continuous}. In fact, DRL algorithms may fail to converge under the interference of improper parameters setting. Nevertheless, the proposed FL-DDPG algorithm is capable to converge when a few constraints are met. If the learning rate, target network update rate and action noise are properly set, FL-DDPG can converge stably, which can be proved by simulation results displayed in Section \ref{section:6}.

The complexity of the FL-DDPG algorithm is largely determined by the size of the neural network employed. Since the local training approach is adopted, each agent trains the neural network by itself, so the complexity of each agent can be denoted as $\zeta^u$ can be calculated independently and the total complexity of the multi-agent system is $\zeta=\zeta^G+\sum\limits_{u=0}^{U}{\zeta^u} $, where $\zeta^G$ represents the complexity caused by the updating and downloading parameters of the globe model.

The action selection for each step is the responsibility of the actor network $\omega^Q$, and we denote that the number of nodes in the actor network as $\omega^Q_\text{n}$ for normalized nodes, $\omega^Q_\text{r}$ for relu nodes and $\omega^Q_\text{t}$ for 'tanh' nodes. Thus, the calculations complexity caused by the node computation is $ 5 \cdot \omega^Q_\text{n} + \omega^Q_\text{r} + 6 \cdot \omega^Q_\text{t}$ as suggested in \cite{qiu2019deep}. Further, assuming the actor network has $I$ layers in total and each layer $i$ has $\|\omega^Q_i\|$ nodes, the complexity required to propagate values between neural nodes and adding bias can be calculated as $\sum \limits_{i=0}^{I}\|\omega^Q_i\|\cdot \|\omega^Q_{i+1}\|$. Then the complexity of actor network for a single step is $\zeta_{\omega^Q}=5 \cdot \omega^Q_\text{n} + \omega^Q_\text{r} + 6 \cdot \omega^Q_\text{t}+ \sum \limits_{i=0}^{I}\|\omega^Q_i|\cdot \|\omega^Q_{i+1}\| $. If we apply the same assumption to the critic network $\mu$, since the critic network has to train $e$ samples at each step, with the same calculation method, the complexity of the critic network is $\zeta_{\omega^\mu}= e \cdot (5 \cdot \omega^\mu_\text{n} + \omega^\mu_\text{r} + 6 \cdot \omega^\mu_\text{t}+ \sum \limits_{i=0}^{I}\|\omega^\mu_i|\cdot \|\omega^\mu_{i+1}\|) $. Then, for the proposed scenario, which has $t$ steps per episode, for a single agent the total complexity can be calculate as $\zeta_u = \mathcal{E} \cdot t \cdot (\zeta_{\omega^Q}+\zeta_{\omega^\mu})$, where $\mathcal{E}$ represents the episode number. On the other hand, the complexity caused by the globe model is $2\cdot\mathcal{E}/F_G \cdot \|\omega^Q\|+\|\omega^\mu\|$, where $F_G $ represents number of episodes interval of global model update, which is negligible compared to the local model training. Therefore, the total complexity can be express as $\zeta=\sum\limits_{u=0}^{U}{\zeta^u_{\omega_Q} + \zeta^u_{\omega_\mu}}$.

\section{Numerical Results and Analysis}\label{section:6}

Section VI aims to exhibit numerical results of the FL-DDPG optimized mobile RIS system. In the simulation, we assume that each cell serves four users and these users are partitioned into two clusters. Each user makes a random movement on the horizontal plane at every time slot, the moving distance conforms to the Rayleigh distribution and the direction following the uniform distribution.
The building structure of each cell is assumed to be the same, and the global model update frequency for the FL is 20 episodes. As for the agent, Adam optimizers are employed for the neural network training and the proper learning rate range is $5\times10^{-4}$ to $10^{-5}$ according to our simulation. The initial action noise scale is set as 0.4.  The rest of the default parameters have been given in Table \ref{SP}.

\begin{table}[t!]

 \caption{Simulation Parameters}\label{SP}
 \centering
 \footnotesize
 \renewcommand\arraystretch{1.5}
 \begin{tabular}{|c|c|c|c|c|c|}
  \hline
  Parameter & Description & Value &   Parameter & Description & Value \\
  \hline
  $f_\text{c}$ & carrier frequency & 2GHz & $K$ & number of users & 4  \\
  \hline
  $B^u_k$ & bandwidth  & 1 MHz & $P_{\text{max}k}^u$ & maximum transmitting power & 20 dBm \\
  \hline
  $V_\text{max}$ & maximum speed of RIS & 0.5 m/s & $\lambda$ & QoS penalty coefficient & 2 \\
  \hline
  $y_\text{max}$ & room length & 20 m & $x_\text{max}$ & room width & 15 m \\
  \hline
  $R_\text{QoS}$& QoS require & 10 kb/s & $\sigma$& noise power density & -30 dBm/MHz \\
  \hline
  $\alpha$ & learning rate & $3\times10^{-4}$ & $\gamma$ & discount factor & 1\\
  \hline
  $e$ & batch size & 64 samples & $\tau$ & target update rate & 0.002 \\
  \hline

 \end{tabular}
\end{table}

\begin{figure}[t!] 
\centering
\includegraphics[width=0.5\textwidth]{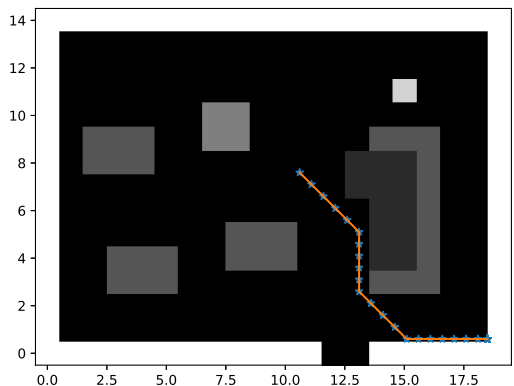} 
\caption{Optimized path for the mobile RIS} 
\label{Fig.path} 
\end{figure}

Fig. \ref{Fig.path} exhibits a trajectory example of the mobile RIS derived from the proposed DDPG algorithm. In this figure, the orange curve records the trajectory of the mobile RIS and the blue stars represent the position where the robot stops at each discrete time slot, which is also the RIS position that is input into the neural network as a part of the state information. The mobile RIS is initially placed in the middle area of the office, and it moves to a corner gradually so that provides LoS cover for the large area blocked by the sofa. The gray and white blocks correspond to the furniture and walls of different heights. It can be observed that the derived path avoids obstacles and the data rate gain for the flexible deployment will be discussed later.

\begin{figure}
  \begin{minipage}[t]{0.5\linewidth}
    \centering
    \includegraphics[scale=0.5]{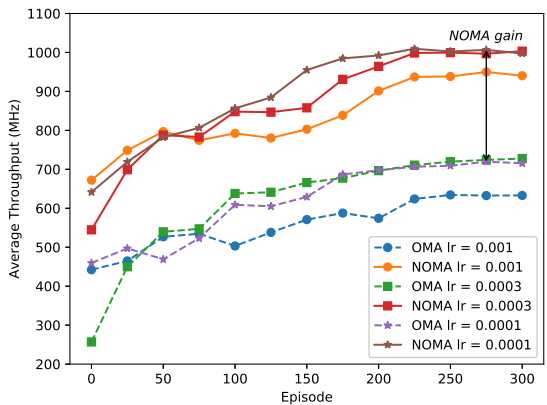}
    \caption {Mobile RIS performance with different learning rates}
    \label{Fig.LR}
  \end{minipage}%
  \begin{minipage}[t]{0.5\linewidth}
    \centering
    \includegraphics[scale=0.5]{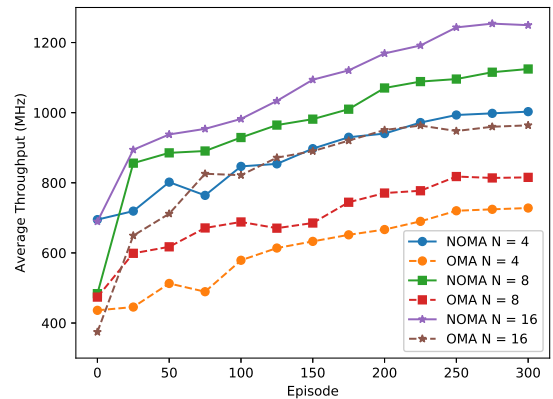}
    \caption {Mobile RIS performance with different reflection element numbers}
    \label{Fig.N}
  \end{minipage}
\end{figure}

Fig. \ref{Fig.LR} demonstrates the training performance of the DDPG algorithm in a single cell. It can be observed that the average throughput of the system increases steadily over the training episodes and gradually flattens out in the late stage of training, which proves that the algorithm has stable convergence within a proper learning rate range. The throughput of well trained agents indicates that an inappropriately large learning rate can result in a debuff in optimization performance. For example, when the learning rate is 0.001, the throughput suffers a decrease of approximately 7\% compared to the other two learning rates. Moreover, we can observe a significant NOMA gain, which is around 42\% compared to the OMA scheme under the same conditions.

The impact of the RIS reflecting elements number on system performance is investigated in Fig. \ref{Fig.N}. Logically, a larger amount of reflection elements can enhance the propagation to a superior extent and obtain further power gain. It can be observed that with the enhancement of 16 reflection elements, the OMA scheme obtain a data rate equivalent to the NOMA scheme with 4 reflection elements. Meanwhile, the stable convergence of results indicates although the different values of reflecting elements number $N$ cause tremendous dimensional differences of the input state, by correspondingly adjusting the size of the neural network, the proposed algorithm can serve RIS with different specifications.

\begin{figure}
  \begin{minipage}[t]{0.5\linewidth}
    \centering
    \includegraphics[scale=0.5]{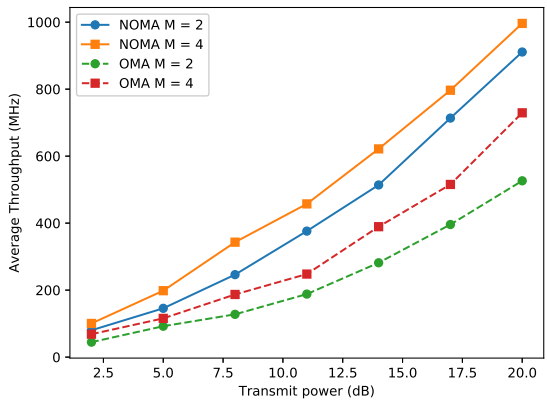}
    \caption {Achievable sum rate versus AP transmit power}
    \label{Fig.M}
  \end{minipage}%
  \begin{minipage}[t]{0.5\linewidth}
    \centering
    \includegraphics[scale=0.5]{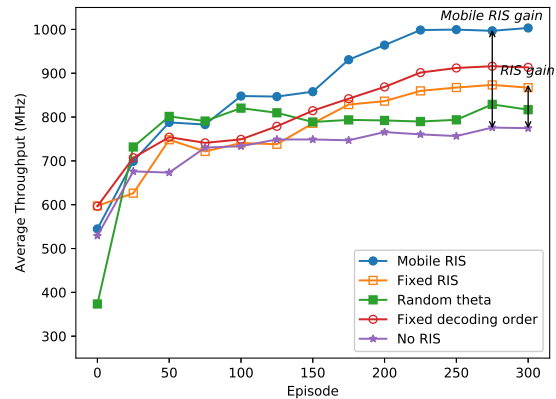}
    \caption {Date rate gain of each component in mobile RIS enhanced networks}
    \label{Fig.Gain}
  \end{minipage}
\end{figure}

We plot the throughput curve versus the transmit power in Fig. \ref{Fig.M} and display both OMA and NOMA cases where the number of antennas $M$ is 2 or 4. The data rate gain of the 4 antennas case is approximately 11.6\% on average, compared to the case of double antennas. The NOMA gain is higher with the growth of the transmission power, the reason is when the transmission power is low, the weaker users are not likely to meet the QoS requirements and need to be allocated more power to ensure the fairness. Although this fairness-dominated power allocation scheme results in a reduction in data rate gain, the NOMA scheme still achieves a noticeable gain in the case of small transmit power.

In order to determine the gain of the maneuver deployment and each other component in the mobile RIS model, we plot Fig. \ref{Fig.Gain} to show the throughput of the proposed model and benchmarks. First of all, the dynamic decoding order achieves a gain of 10.2\% compared to the pre-settled static decoding order. By observing the curve, it can be found that in this case the flexibly deployed RIS obtains an additional 15.1\% data rate gain compared to the fixed RIS scheme, where the RIS is settled at the start position in Fig. \ref{Fig.path}.  It is worth noting that the performance improvement provided by the mobile RIS even exceeds the performance difference between the fixed RIS model and no RIS engaged network, which indicates the superiority of the mobile RIS framework is substantial and puissant. In addition, in contrast to the fixed RIS model, the mobile RIS has compelling compatibility for various user distributions. In order to investigate the effect of the phase shift optimization, we employ a RIS with random phase shifts as another benchmark. It is undeniable that the RIS with random phase shifts also leads to a diminutive gain compared to the no RIS mode, but it is far inferior than the DRL optimized case. Meanwhile, the curve behaves unevenness even in the final episodes since the phase is not controlled by the agent.

\begin{figure}
  \begin{minipage}[t]{0.5\linewidth}
    \centering
    \includegraphics[scale=0.5]{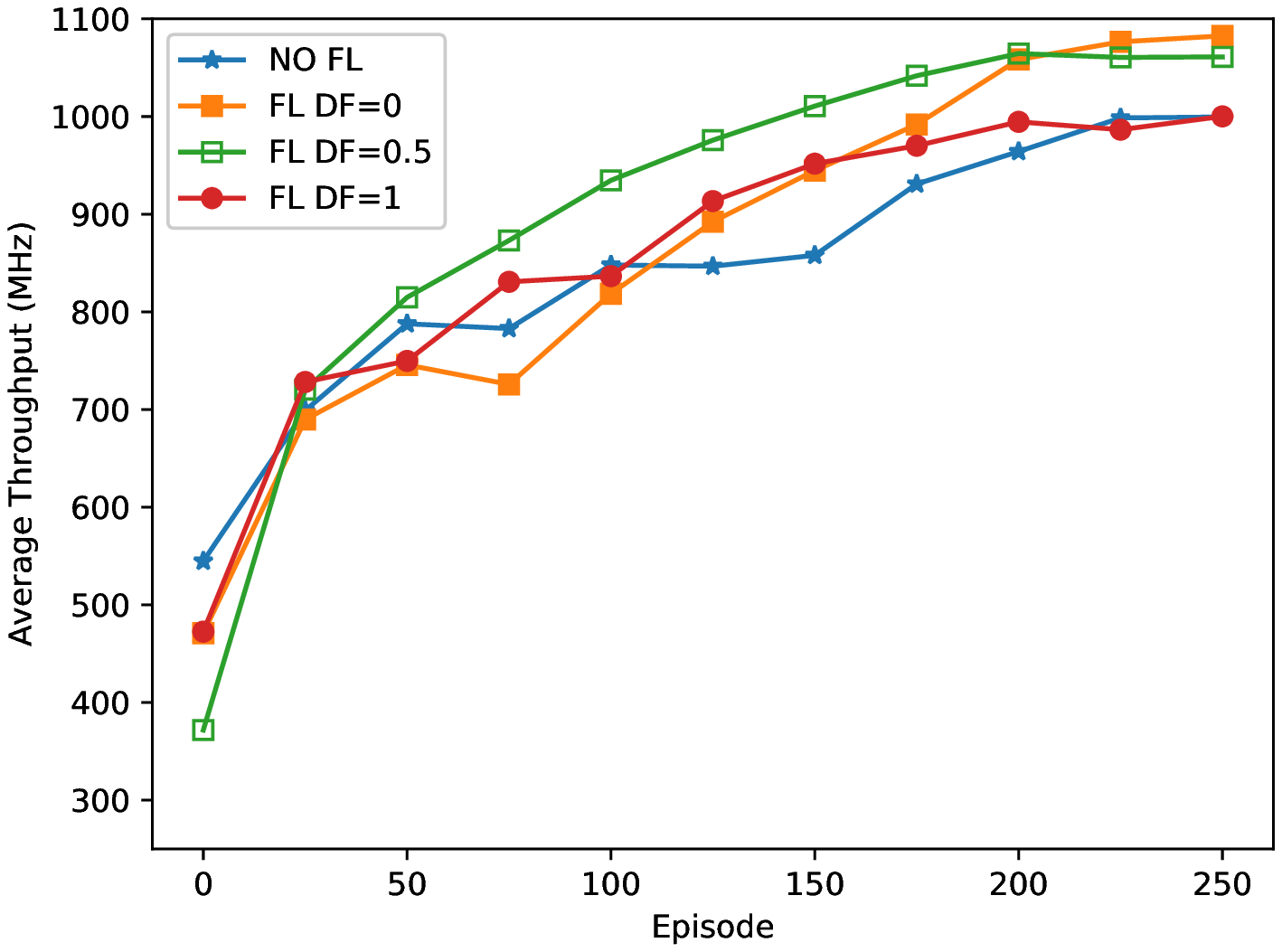}
    \caption{\small The performance of federated learning}
    \label{Fig.FLDF}
  \end{minipage}%
  \begin{minipage}[t]{0.5\linewidth}
    \centering
    \includegraphics[scale=0.5]{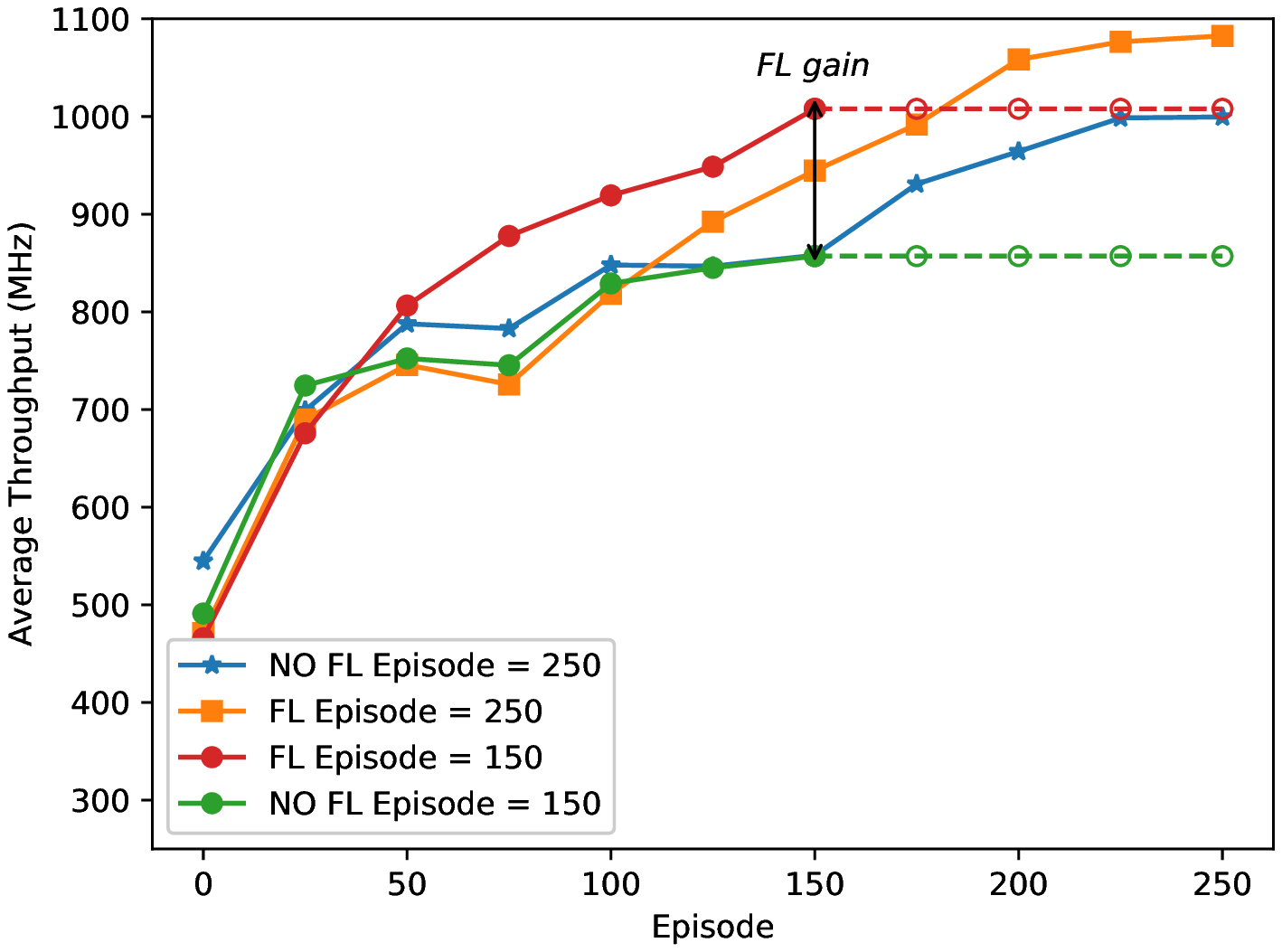}
    \caption{\small Training effect with/without federated learning}
    \label{Fig.FLTRAIN}
  \end{minipage}
\end{figure}

Fig. \ref{Fig.FLDF} shows the impact of environmental differences on the performance of federated learning, where the difference factor (DF) represents the correlation of the fading characteristic in different cells. DF is 0 means that the cells have the same channel characteristics. Obviously, FL achieved the optative performance in this case, since agents are in the same environments so that the model update has the highest efficiency. It is worth noting that even though the rooms have similar architectural structures, they have different fading characteristics due to the difference in decoration and surface materials. Therefore, we investigate the cases that cells with propagation differences, and DF = 1 suggests that the propagation characteristics of each cell are completely independent. It can be observed that even in the case of DF=1, FL-enhanced DRL still has stable convergence, and is capable to achieve a matched average sum rate to the single-cell case.

We intend to plot Fig. \ref{Fig.FLTRAIN} to reveal the gain of FL at different training maturities. It can be found at first that the introduction of FL can effectively save training time. With the aid of FL, agents only spend 150 episodes of training to achieve an equal performance that the single-cell scheme needs 250 episodes, which supports the statement in \textbf{Remark \ref{rmark.FL}}. Since the DRL approaches train agents by replaying the obtained experiences, more diverse and richer experience of transitions obtained by FL makes the agents' decision-making wiser. It is undeniable that spending infinite training episodes can enable all states to be explored, so that the agents can converge to the same optimal level. However, in practice, under the condition that the training time is limited, FL has a significant training advantage compared to the mode without FL.

\section{Conclusions}\label{section:7}

This paper has proposed a NOMA enhanced wireless network model with the aid of mobile RISs that can provide NOMA craved channel conditions and improve channel quality for users. In order to optimize deployments and phase shifts of RISs and the corresponding power allocation for users, an FL enhanced DDPG algorithm has been proposed, which has preponderant performance under the same training extend compared to the independent DLR scheme since the engagement of FL lead to more sufficient exploration and experience exchange for agents. Simulation results proved that 1) Compared to the scenario without RIS, mobile RISs are capable to provide around 30.1\% data rate gain that significantly exceeds the gain of the fixed RISs paradigm, which is 12.4\%; 2) The NOMA scheme, where the proposed dynamic decoding identification method is applied, outperforms the OMA scheme by obtaining approximately 42\% gain in terms of the sum rate.; 3) The FL enhanced DDPG algorithm has stable convergence while the parameters are within an appropriate range and the participation of the federated learning can considerably reduce the training time of the DDPG agents or improve the training effect under a limited equal training process.


\renewcommand{\baselinestretch}{1.05}
\bibliography{RIS_Robotref_abb}
\bibliographystyle{IEEEtran}

\end{document}